\documentclass[twocolumn]{revtex4}

\usepackage{float}
\usepackage{graphicx}
\usepackage{color}
\usepackage{amssymb}

\newcommand{\msd}{\langle x^2 (t)\rangle}

\newcommand{\DD}{{\cal D}}

\newcommand{\kd}{k_D}
\newcommand{\kon}{k_{\rm on}}
\newcommand{\koff}{k_{\rm off}}
\newcommand{\toff}{\tau_{\rm off}}
\newcommand{\tc}{\tau_{\rm coll}}

\newcommand {\be} {\begin{equation}}
\newcommand {\ee} {\end{equation}}
\newcommand {\bea} {\begin{eqnarray}}
\newcommand {\eea} {\end{eqnarray}}

\begin{document}

%%%%%%%%%%%%%%%%%%%%%%%%%%%%%%%%%%%%%%%%%%%%%%%%%%%%%%%%%%%%%%%%%%%%%
\title{%Diffusion of a Tracer Particle in a Crowded Nearly One Dimensional System \\
Many--body effects in  tracer particle diffusion  with applications for single-protein dynamics on DNA}

\author{Sebastian Ahlberg}
\affiliation{Integrated Science Lab, Department of Physics, Ume{\aa}
University, SE-901 87 Ume{\aa}, Sweden}

\author{Tobias Ambj\"ornsson}
\affiliation{Department of Astronomy and Theoretical Physics, Lund University,
  S\"olvegatan 14A, SE-223 62 Lund, Sweden}

\author{Ludvig Lizana}
\affiliation{Integrated Science Lab, Department of Physics, Ume{\aa}
University, SE-901 87 Ume{\aa}, Sweden }
\email{ludvig.lizana@physics.umu.se}

\date{\today}

%%%%%%%%%%%%%%%%%%%%%%%%%%%%%%%%%%%%%%%%%%%%%%%%%%%%%%%
%
% 							A B S T R A C T
%
%%%%%%%%%%%%%%%%%%%%%%%%%%%%%%%%%%%%%%%%%%%%%%%%%%%%%%%

\begin{abstract}

 30\% of the DNA in {\it E. coli} bacteria is covered by proteins. Such high degree of crowding affect the dynamics of  generic biological processes (e.g.  gene regulation, DNA repair, protein diffusion etc.) in ways that are not yet fully understood. In this paper, we theoretically address the diffusion constant of a tracer particle in a one dimensional system surrounded by impenetrable crowder particles. While the tracer particle always stays on the lattice, crowder particles may unbind to a surrounding bulk and rebind at another or the same location. In this scenario we determine how the long time diffusion constant $\DD$ (after many unbinding events) depends on (i) the unbinding rate of crowder particles $\koff$, and (ii) crowder particle line density $\rho$, from simulations (Gillespie algorithm) and analytical calculations.
%
%Using extensive simulations (Gillespie algorithm) as well as analytical calculations, we determine how the long time diffusion constant $\DD$ (after many unbinding events) depends on (i) the unbinding rate of crowder particles $\koff$, and (ii) crowder particle line density $\rho$. 
 %
 For small $\koff$, we find  $\DD\sim \koff/\rho^2$  when crowder particles are immobile on the line, and $\DD\sim \sqrt{D\koff}/\rho$  when they are diffusing; $D$ is the free particle diffusion constant. For large $\koff$, we find agreement with  mean-field results which do not depend on $\koff$.  From literature values of $\koff$ and $D$, we show that the small $\koff$-limit is relevant for {\it in vivo} protein diffusion on a crowded DNA. Our results applies to single-molecule tracking experiments.
\end{abstract}
%\pacs{}

\maketitle

%%%%%%%%%%%%%%%%%%%%%%%%%%%%%%%%%%%%%%%%%%%%%%%%%%%%%%%
%
%							I N T R O D U C T I O N
%
%%%%%%%%%%%%%%%%%%%%%%%%%%%%%%%%%%%%%%%%%%%%%%%%%%%%%%%
\section{Introduction}

Few doubt that molecular crowding has severe consequences for dynamical processes \cite{zhou2008macromolecular}.  Interesting examples are living cells where macromolecular concentrations are large. Take the {\it E. coli} bacterium as an example. There, the concentration of proteins and RNA  is about 300--400 mg/ml \cite{zimmerman1991estimation} which is 30--40 times larger  than common test tube conditions \cite{ellis2001macromolecular}. There is overwhelming evidence that this level of crowding influences important biological processes such as gene regulation \cite{li2009effects}, enzymatic activity \cite{nakano2009facilitation}, protein folding \cite{zhou2004protein,martin2002requirement}, and diffusion of macromolecules \cite{banks2005anomalous, hofling2013anomalous}.  In order to get a complete picture of the {\it in vivo}  dynamics  we must increase our understanding of the role of crowding, and recent experimental developments  provide the means to do it.

In recent years, researchers have beat the diffraction limit and turned optical microscopy into 'nanoscopy'.  Today's microscopy methods (e.g. STED, STORM and FIONA) \cite{knorpp2090method}  does not only allow us to image nanometer-sized biological structures, but recent improvements \cite{persson2013single,heller2013sted} also permit tracking fluorescently labelled proteins at the biologically relevant millisecond-scale. This  is anticipated to shed new light on biological processes, as well as increase our understanding of particle transport in engineered nano-fluidic systems \cite{karlsson2001molecular, dekker2007solid}.  In order to properly interpret those type of experiments {\it in vivo}, we need new theoretical and computational models  in terms of physical properties of the intracellular space, the cytoplasm.

The cytoplasm is cramped with macromolecules and we are interested how this influences diffusion-controlled processes,  a key component in many cellular functions (e.g. gene regulation).   While our results are new, aspects of this problem has been studied theoretically  before.
For example, \cite{ando2010crowding, dix2008crowding, mcguffee2010diffusion, netz1997computer} investigate diffusion in the three dimensional cytoplasm and  in gels, whereas  \cite{weiss2004anomalous, ghosh2014non, saxton1996anomalous, szymanski2009elucidating} focus on the sub-diffusive motion seen in single-molecule experiments. Crowding is also important for DNA search processes where a searcher combines one  and three dimensional diffusion  to quickly find its target, so called facilitated diffusion. Facilitated diffusion under crowding is addressed in  \cite{marcovitz2013obstacles, li2009effects}  which resemble this paper but we ask different questions: we calculate the diffusion constant of a tracer particle in terms of  key properties of surrounding crowder particles rather than focusing on mean target finding times.

\begin{figure}
\includegraphics[width = 0.8\columnwidth]{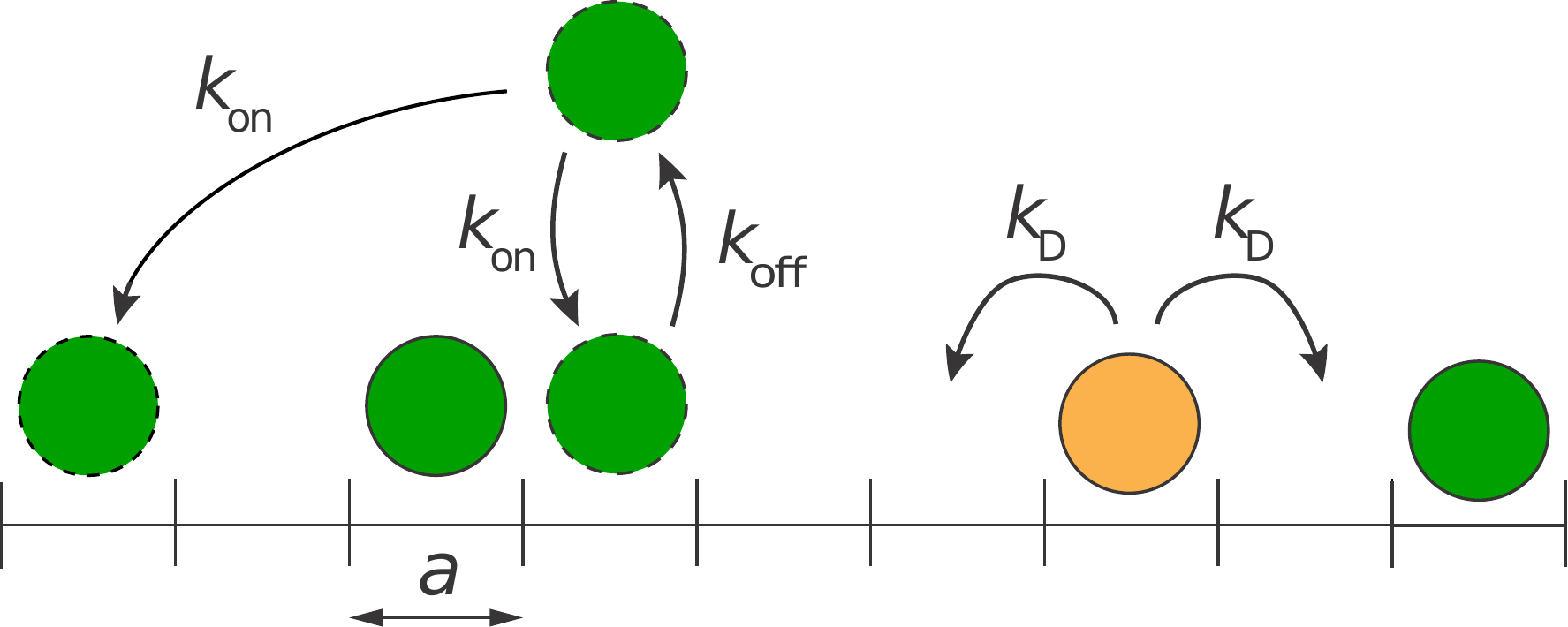}
\caption{Schematic illustration of our model. All particles are diffusing with
  rate $\kd$ on a one dimensional lattice with lattice spacing $a$.  The crowder particles (green) may also unbind and rebind
 to a random, or  the same, lattice site  with rates $\koff$ and $\kon$, respectively. The tagged particle (orange) cannot
  leave the line ($\kon=\koff=0$). }
\label{fig:model}
\end{figure}

%In this paper we focus our attention on a simple one dimensional crowded system which exchange particles with a  surrounding bulk. 
Much  inspiration to this work comes from DNA binding proteins. Of particular interest  is  repair proteins (MutS and homologs) whose residence time on the DNA can be very long ($\sim 10$ min \cite{cho2012atp}). We are also inspired by transcription factors, the family of gene regulatory proteins. The yeast regulatory proteins LexA and Gal4 can stay bound to their  regulatory sites for several minutes  {\it in vitro} (LexA $\sim$ 5 min and Gal4 $>$ 30 min) \cite{luo2014nucleosomes}, but surprisingly, this number can be reduced  up to 1000 times {\it in vivo}. Both classes of  proteins have the ability to diffuse along the DNA, unbind to the three dimensional intracellular space,  diffuse in space, and rebind to the DNA. We are  interested in how the dynamics of those proteins change under crowding.

In order to better understand the role of crowding, we introduce a theoretical model  where particles diffuse on a one dimensional lattice where two particles cannot occupy the same site (Fig. \ref{fig:model} ). They diffuse with rate $\kd$ (same for all particles) and  may unbind  and rebind to the lattice with rates $\koff$ and $\kon$, respectively. These rates are tuned so that the average particle line density is constant at 10-20\% which is not too far from {\it in vivo} conditions (30\% of the DNA in {\it E. coli} is  covered by proteins). The  unbinding rate for the tracer particle is set to zero similar to the long-lived protein-DNA complexes described above. Now we ask: \\

\noindent
{\it What is the long time diffusion constant of a tracer particle in such a crowded  quasi one dimensional system? }\\

\noindent
We answer this question numerically using stochastic simulations (Gillespie algorithm), corroborated with analytical results. The main results are  Figs. \ref{fig:mobility_small}--\ref{fig:mobility_interpol} where we show how the diffusion constant changes as a function of our main parameter $\koff$.
%, for different binding/unbinding modes (rebinding to the same or random lattice site). 
%Since we focus on the long-time diffusion constant we do not pay attention to  possible transients on the way.  
Those results are applicable to single molecule tracking experiments \cite{gorman2008visualizing}. 
%First-passage properties of this system was characterised in a previous publication \cite{forsling2014non}.

The unbinding rate  $\koff$  interpolates between two well studied limits. (i) When $\koff$ is large (compared to $\kd$), the tracer's mobility is only weakly lowered and  diffuses close to as if it was free. (ii) When $\koff\rightarrow 0$, the particles diffuse with unchanged order  in a single file. Single-file diffusion is well studied \cite{harris1965diffusion, levitt1973dynamics,percus1974anomalous, kollmann2003single, lizana2008single, barkai2009theory, lomholt2011dissimilar} where  the most famous result is that the  mean squared displacement of a tracer tracer particle is proportional to $\sqrt t$ rather $t$ ($t$ is time) which signatures non-markovian dynamics.

This paper is organised as follows. In Sec. II, we outline briefly the details of our model. Before showing the results in Sec. IV, we provide  analytical estimates of the diffusion constant in Sec. III, based on a theoretical calculation found in Appendix \ref{appendix_1}.  In Sec. III we also briefly review the dynamics of the model at short, intermediate and long times. We close by a few concluding remarks in Sec. V.

%%%%%%%%%%%%%%%%%%%%%%%%%%%%%%%%%%%%%%%%%%%%%%%%%%%%%%%
%
%							T H E  M O D E L
%
%%%%%%%%%%%%%%%%%%%%%%%%%%%%%%%%%%%%%%%%%%%%%%%%%%%%%%%
\section{The model}\label{sec:model}

Our model has been used and explained elsewhere \cite{forsling2014non}, but for completeness we summarise it briefly below.
Consider a one dimensional lattice on which crowder particles (assumed identical) and the tracer
particle diffuse (Fig. \ref{fig:model}). The crowder particles can diffuse, unbind and rebind to the lattice.  Rebinding occurs in two ways. Either to a random unoccupied lattice site  (chosen with uniform probability), or to the exact same location. Both rebinding modes  has been used to model transcription factor dynamics on DNA \cite{kolomeisky2011physics,benichou2011intermittent}, and we will therefore consider both. 
%The case when the particle rebinds to the same location is also more amenable to theoretical treatment, especially when crowder particles are not allowed to diffuse.
%
The lattice constant is denoted $a$, and the diffusion rate $\kd$ is assumed equal in both
directions and for all particles. Double occupancy is forbidden and a particle cannot overtake a flanking neighbor (single-file condition). Binding and unbinding dynamics of crowders are characterised by the rates $\kon$ and $\koff$, which are chosen such that the particle line density is in equilibrium with the bulk, thereby keeping the average filling fraction is constant. In our simulations we keep it at  10--20\%. We implemented the model  using the Gillespie algorithm. See Appendix \ref{sec:numerics} for details.

%We implemented the model described above using the Gillespie algorithm;
%details are deferred to Appendix \ref{sec:numerics}. Briefly, in a simulation we
%initially place the particles randomly (thermal initial condition) where the
%initial position of the tracer particle is 70 sites away from the target site. Also, due to recent
%interest in non-thermal initial conditions in single-file diffusion
%\cite{lizana2014single,leibovich2013everlasting} we in addition investigate
%the case when the particles are placed equidistantly for $\koff=0$.
%We run the simulation until the tagged particle hits the absorbing target and
%make a record of the absorption time. From many such runs we then determine
%the ensemble averaged FPTD (normalized histogram of absorption times) as a
%function of our key parameter $\koff$.  In this way we interpolate between the
%single-file regime ($\koff \ll \kd$) to the unobstructed single particle
%regime ($\koff \gg \kd$).

%%%%%%%%%%%%%%%%%%%%%%%%%%%%%%%%%%%%%%%%%%%%%%%%%%%%%%%
%
%							A N A L Y T I C S
%
%%%%%%%%%%%%%%%%%%%%%%%%%%%%%%%%%%%%%%%%%%%%%%%%%%%%%%%
\section{Analytical estimates for the long time diffusion constant}\label{sec:analyt}

Here we provide analytical estimates to corroborate and better understand the numerical
results in the next section. We are mainly interested in the long time diffusion $\DD$ constant for the tracer particle, defined as
\be \label{eq:long_time_D}
\msd \simeq 2 \DD t
\ee
where $\msd$ is the ensemble averaged mean squared displacement (MSD), and $t$ is time. Notably, $\DD$ is in general not equal to the bare, or free particle, diffusion constant 
\be \label{eq:D_bare}
D=a^2k_D.
\ee
It is a non-trivial function of $\koff$ and $\rho$. To better understand  what we mean by long time, we describe in subsection \ref{sec:transient_dyn}   the   dynamics leading up to Eq. (\ref{eq:long_time_D}). But first we summarise our main analytical findings from Appendix  \ref{appendix_1} which we in  Sec. \ref{sec:results} compare  to  simulations.

\subsection{Long time, small $\koff$ behaviour}

%When $\kd\ll\koff$ the tracer particle has to wait for the neighbouring particles to diffuse away or detach from the lattice. 
To simplify matters, we start by assuming that the crowder particles sit equidistantly on the line with density $\rho$,  unable to diffuse ($\kd = 0$), and rebind to the site from which they unbound. In this situation, the tracer moves back and fourth between its flanking neighbours and can only move past them if one of them unbinds. The average time until this happens is proportional to $1/\koff$.
In point of view of the tracer this process is a random walk on an effective, or coarse grained, lattice with spacing  and jump rate proportional to $1/\rho$ and $\koff$, respectively. From this we expect that $\DD \sim \koff/\rho^2$, and  a more elaborate calculation shows that 
\be\label{eq:D_gate}
\DD \simeq \frac {3 (1-a\rho)\koff} {4 \rho^2 }.
\ee

When crowder particles rebind to a random location rather than to the same site, the distance between two neighbouring particles  fluctuate even though the average  density is fixed. This leads to a larger effective lattice spacing, and  a larger $\DD$ compared to Eq. (\ref{eq:D_gate}):
\be\label{eq:D_rnd}
\DD \simeq \frac{(2-a\rho)(1-a\rho)}{4\rho^2} \koff.
\ee

When crowder particles also diffuse ($\kd \neq 0$) the distance between nearest--neighbours becomes difficult to define.  We estimate the coarse grained lattice constant as the length the tracer particle explores during a time intervall proportional to $1/\koff$. This leads to 
\be\label{eq:D_sfd}
\DD \simeq \frac{(1-a\rho)^{3/2}}{\rho} \sqrt{\frac{D\koff}{2\pi}},
\ee
which has different  $\koff$-scaling than before.
%Admittedly, this case is more involved  and we push our theory to the boundary of its applicability. 
%For instance, we know that a single-file diffusion process belongs to the class of fractional Brownian motion (with Hurst exponent 1/4), whereas CTRW in general is non-Gaussian. 
%
Equations  (\ref{eq:D_gate})--(\ref{eq:D_sfd})  constitute our main analytical results.
%, and they are compared to numerical simulations in Sec. \ref{sec:results}.

\subsection{Long time, large  $\koff$ behaviour}
When $\kd\gtrsim\koff$,  crowder particles frequently unbind and rebind to the lattice and the no-passing condition is effectively violated. But, crowder particles still hinder the tracer thereby decreasing the diffusion rate. Imagine that the jump rate for a single particle to a neighbouring site on an otherwise empty lattice is $\kd$, or $D=a^2\kd$. Then, when crowder particles are around, some of the jumps are canceled because the target lattice site may be occupied. In that situation the jump rate is reduced by the probability that the target lattice is unoccupied. For very large $\koff$ this probability is simply $1-a\rho$, therefore
\be\label{eq:D_large}
\DD \simeq D(1-\rho a)
\ee
This mean field result  has been obtained before \cite{haus1987diffusion, nakazato1980site, ghosh2014non}, and as $\koff/\kd$ is close to or smaller than unity, corrections to this formula becomes prominent (see Fig. \ref{fig:mobility_large}).

\subsection{Interpolation formula for $\DD$}
Based on the expressions above, we propose a simple formula for $\DD$ valid for all $\koff$: 
\be\label{eq:DD_full}
\frac 1 \DD  = \frac 1 {\DD_{ {\rm small} \ \koff}} + \frac 1 {\DD_{ {\rm large} \ \koff}}
\ee
Here ${\DD_{ {\rm large} \ \koff}}$ is Eq. (\ref{eq:D_large}) whereas ${\DD_{ {\rm small} \ \koff}}$ is one of Eqs. (\ref{eq:D_gate})-(\ref{eq:D_sfd}) depending on the case under study.  Equation (\ref{eq:DD_full}) is appealingly simple  and captures properly the small and large $\koff$ limits, but it should not be viewed as more than a candidate expression for $\DD$. We have not made a systematic attempt to find the best  form of $\DD$, and leave it for future research.

\subsection{How is the long time asymptotics [Eq. (\ref{eq:long_time_D})] approached?}
\label{sec:transient_dyn}

Here we clarify the meaning of  short, intermediate and long times within of our model. To keep the discussion simple, we consider $\koff$, $\kon$, $\kd$, and $\rho$ as constant. See also Fig. \ref{fig:msd_diffusion} which shows $\msd$ as a function of time, where all relevant regimes are present.

At most we have three regimes of different behaviour.  These are separated by  the average residence time   of the crowder particles $\toff$, and the average collision time $\tc$, which is the time it takes for a particle to diffuse across the average nearest neighbour distance $1/\rho$:
\be \label{eq:toff} 
\tc = \frac 1 {\rho^2 D}, \ \ \ \toff = \frac 1 \koff
\ee
Let us assume that there is a clear separation between these timescales and that $\tc\ll \toff$ and that $\kd$ is the fastest rate in the system,  $1/\kd \ll \tc $. In the first regime, $t\ll \tc$, the tracer diffuse as if it was free, since it has not yet collided with its nearest neighbours. This means that the tracer's MSD is $\msd = 2 D t$. In the second regime, $\tc\ll t \ll \toff$, many particle collisions have taken place but particles diffuse with maintained order since they are unable to pass each other. This is the single-file diffusion regime which is characterised by Harris' law $\msd \propto \sqrt{Dt/\rho^2}$ \cite{harris1965diffusion}. Here, memory effects dominate and is the very reason to the sub diffusive behaviour. In the third regime, $t\gg\toff$, particles start unbinding from the lattice which effectively violates the no-passing condition. In this regime we expect diffusive behaviour again $\msd \sim t$, but with a diffusion constant different from $D$,  denoted by $\DD$ [see Eq. (\ref{eq:long_time_D})]. This is the one we wish to calculate, in particular in terms of our key parameter $\koff$. Note that the second regime can be erased completely if we lower $\toff$ such that $\toff \approx \tc$ (or smaller). Similarly, the third regime is absent if unbinding is not allowed, i.e. $\koff=0$ (or $\toff=\infty)$. In most of our simulations, diffusion is the fastest process in the system   which also is the likely scenario a biological cell (see Sec. \ref{sec:conclusion}). To sum up,
\be
\msd \simeq \left\{ 
\begin{array}{lc}
2Dt, 			& t \ll \tc \\
(1-\rho a)/\rho \times \sqrt{4Dt/\pi},  & \tc\ll t \ll \toff \\
2 \DD t,               & t \gg \toff
\end{array}
\right.
\ee

%In the two subsections below we provide analytical estimates for the long time diffusion constant $\DD$ in terms of the unbinding rate $\koff$. First, we start with the small $\koff$ limit ($\koff \ll \kd$) where unbinding is rare. This limit can be mapped onto a CTRW process  on a coarse grained lattice with an exponential waiting time distribution between consecutive jumps. For the opposite limit ($\koff \gtrsim \kd)$, the particles  overtake each other frequently, and the mapping a CTRW process no longer holds. In order to estimate $\DD$ for this case, we modify results from a $d=2,3$ dimensional hard-sphere system. In the end, we will suggest an interpolation formula bridging these limits.

%%%%%%%%%%%%%%%%%%%%%%%%%%%%%%%%%%%%%%%%%%%%%%%%%%%%%%%
%
%							R E S U L T S
%
%%%%%%%%%%%%%%%%%%%%%%%%%%%%%%%%%%%%%%%%%%%%%%%%%%%%%%%

\section{Results}\label{sec:results}

In this section we present  results from stochastic simulations of the model outlined in Sec. \ref{sec:model}, together with of our theoretical findings from Sec. \ref{sec:analyt}. The simulation details can be found in Appendix \ref{sec:numerics}. First, we show  the tracer particle's MSD as a function of time, from which we extract the long time diffusion constant $\DD$. Second, we investigate $\DD$ separately for large and small $\koff$. Finally, we compare our proposed interpolation formula Eq. (\ref{eq:DD_full}) to the full range of $\koff$ values.

\subsection{Dynamics of the model and extraction of the long time diffusion constant}
\begin{figure}
\includegraphics[width = \columnwidth]{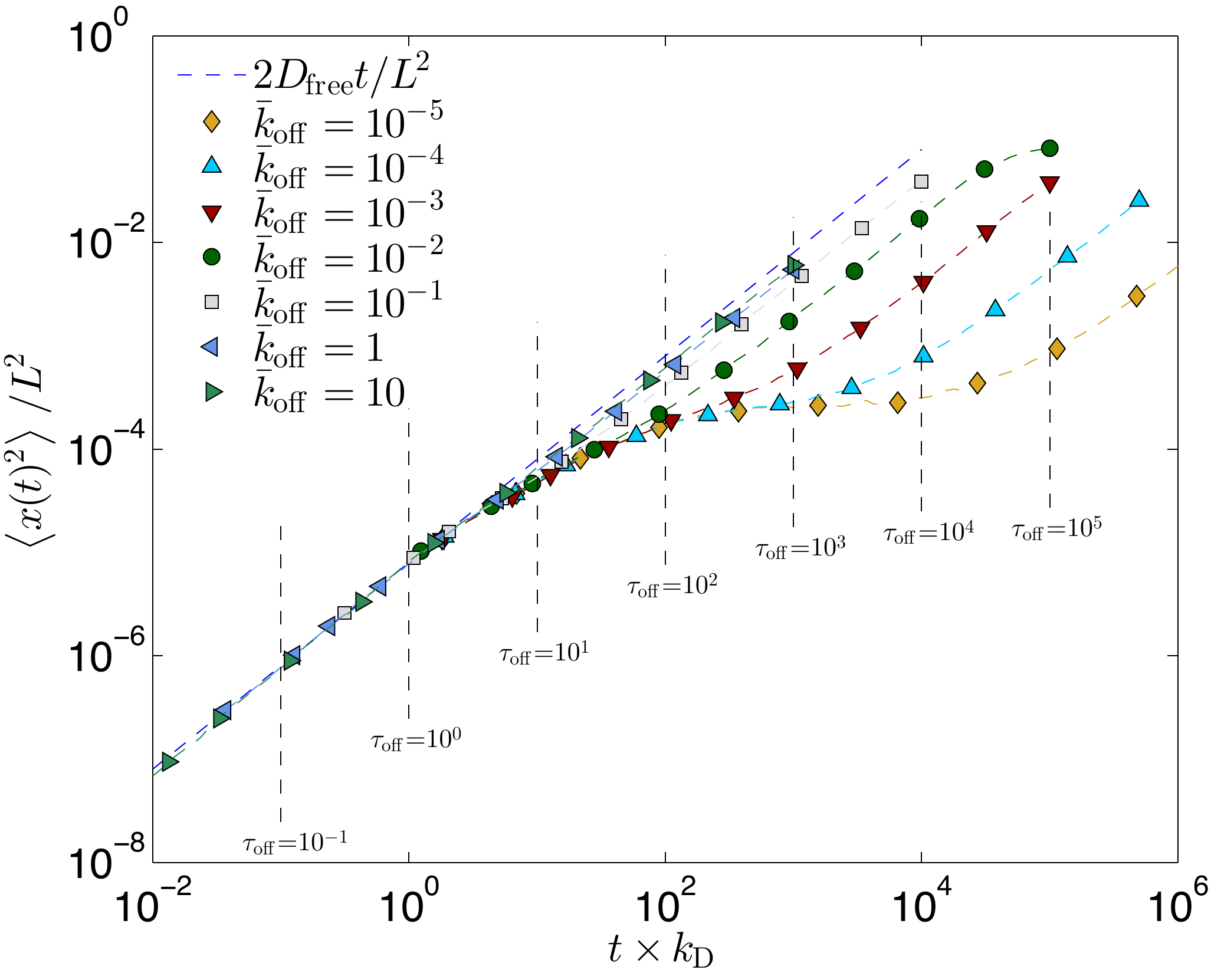}
\includegraphics[width = \columnwidth]{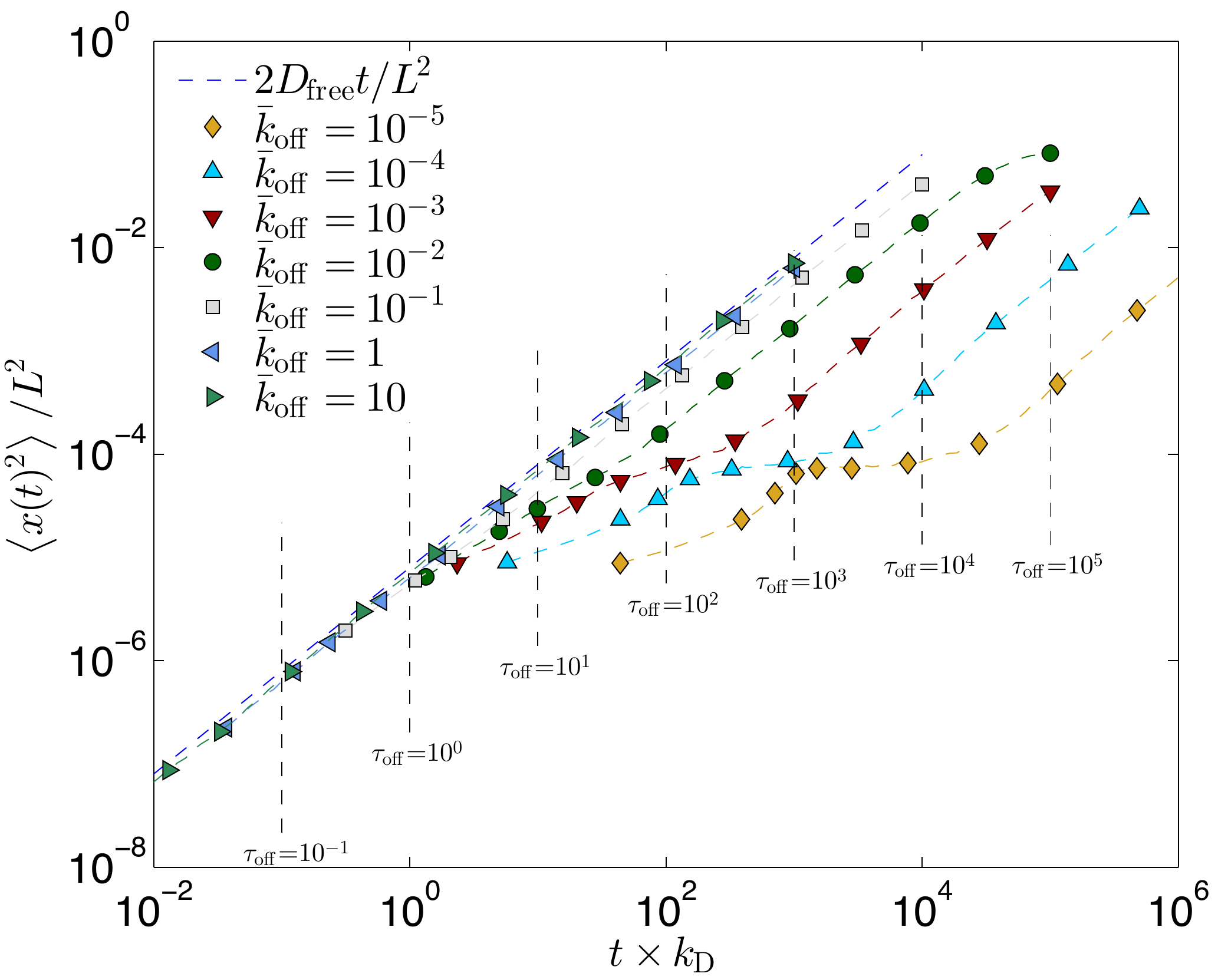}
\caption{Mean squared displacement $\msd$ of the tracer particle as a function from time for different unbinding rates $\koff$ when crowder particles do not move. The crowder particles rebind in two ways: to the same site (upper panel), or to a randomly chosen site (lower panel). For shorthand we put $\bar \koff=\koff/\kd$ and $\bar \toff = \toff \kd$. Simulation details:  lattice constant: $a=1$, tracer particle diffusion rate: $\kd=1$ ($\kd=0$ for crowder particle), filling fraction: $a\rho=0.1$, number of lattice sites: 501 ($L=501a$), number of simulation runs: 9600.}
\label{fig:msd_no_diffusion}
\end{figure}

\begin{figure}
\includegraphics[width = \columnwidth]{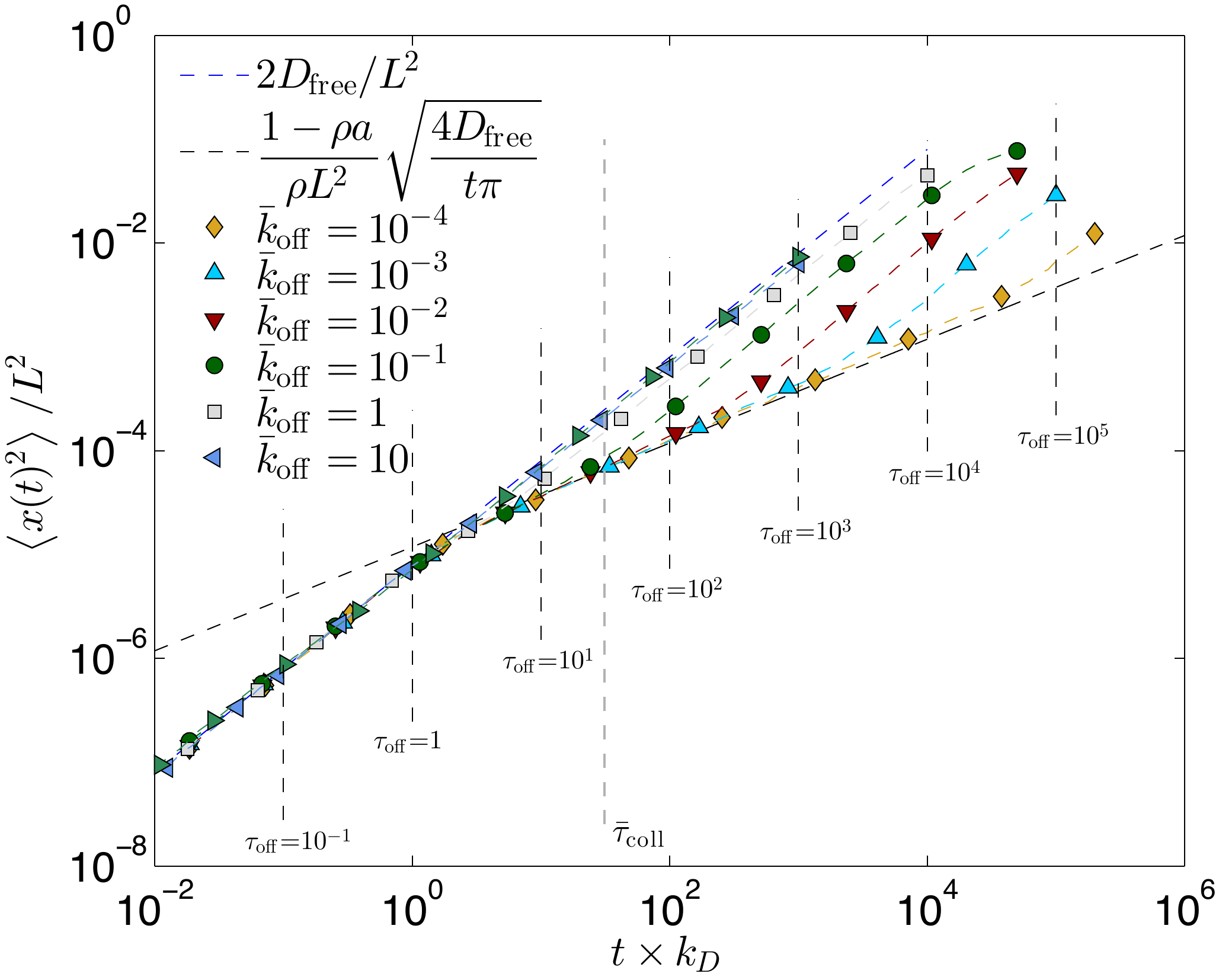}
\caption{Mean squared displacement $\msd$ of the tracer particle as a function from time for different unbinding rates $\koff$ when crowder particles  diffuse. Rebinding occurs to a randomly chosen site. For shorthand we put $\bar \koff=\koff/\kd$,  $\bar \toff = \toff \kd$, and $\bar \tc = \tc\kd$. Simulation details:  lattice constant: $a=1$, diffusion rate: $\kd=1$ (for all particles including the tracer), filling fraction: $a\rho=0.1$, number of lattice sites: 501 ($L=501a$), number of simulation runs: 9600.}
\label{fig:msd_diffusion}
\end{figure}

Figures \ref{fig:msd_no_diffusion} and \ref{fig:msd_diffusion}  show the MSD of the tracer particle as a function of time, for different unbinding rates $\koff$. Symbols represent simulation results. From such plots we extract the long time diffusion constant $\DD$  (see Appendix \ref{sec:linear_fit}) by fitting a straight line for large times starting from $t=\toff$ (short vertical dashed lines).  The results for $\DD$ is shown in Figs. \ref{fig:mobility_small}--\ref{fig:mobility_interpol}, but  first we discuss some of the features of Figs.  \ref{fig:msd_no_diffusion} and \ref{fig:msd_diffusion}.

Figure \ref{fig:msd_no_diffusion} shows the MSD when crowder particles do not diffuse but only unbind and rebind. They rebind either always to the same site (upper panel), or to a randomly chosen site (lower panel). The short time behaviour in both plots is independent of  $\koff$ and is well represented by $2Dt$ (upper dashed dark blue line). The long time behaviour is, however, strongly dependent on $\koff$, which is evident from the broad scattering of curves. The MSD  is still linear in time but the diffusion constant (proportional to the extrapolated intersection with the vertical axis) depends strongly  on $\koff$. The linear regime sets in when $t\approx\toff$  as is seen from the shorter vertical dashed lines. If we increase the particle concentration, the shape of the curves remains the same but the scattering of curves increases, since  $\DD \propto 1/\rho^2$ (small $\koff$) and $\DD \propto 1-a\rho$  (large $\koff$). 
%{\bf Say something about the wiggly part of the curve for large $\koff$??}

In Fig. \ref{fig:msd_diffusion}, crowder particles diffuse and rebind to a randomly chose site. As we lower $\koff$  the separation between $\toff$ and the collision time $\tc$ increases, which means that the single-file regime ($\msd\sim \sqrt t$)  becomes wider. This is simply because crowder particles have not yet started to unbind from the lattice and  therefore diffuse collectively in a single-file. We also see that the MSD curves for long times is less scattered than before, indicating that $\DD$ is less sensitive to $\koff$. This agrees  with our theoretical prediction where $\DD\propto \sqrt{\koff}$ compared to $\DD\propto \koff$ when crowder particles stand still. %In other words, the tracer particle explores a larger part of the system for a given time $t$.

\subsection{Small $\koff$ behaviour}
\begin{figure}
\includegraphics[width = 0.9\columnwidth]{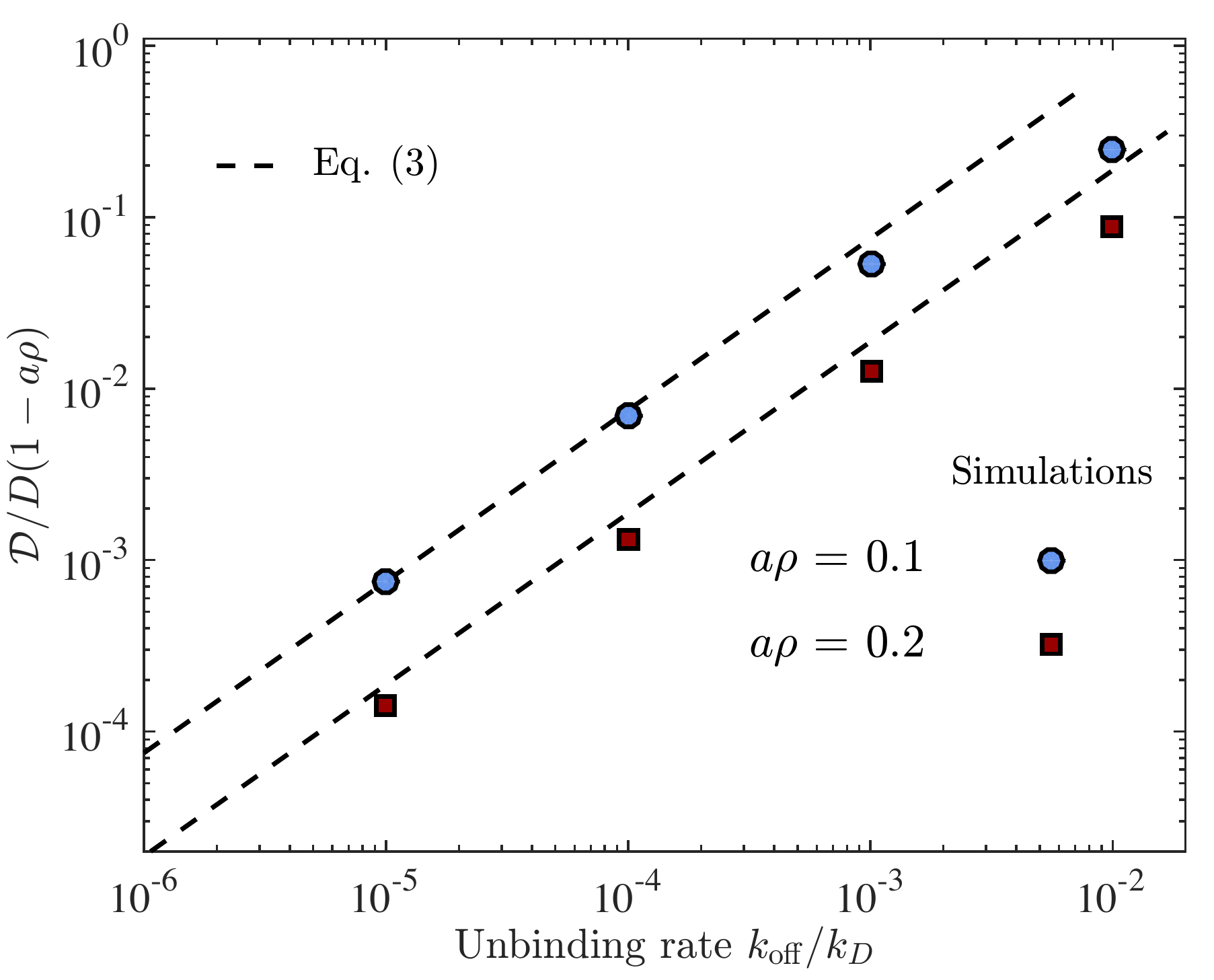}
\includegraphics[width = 0.9\columnwidth]{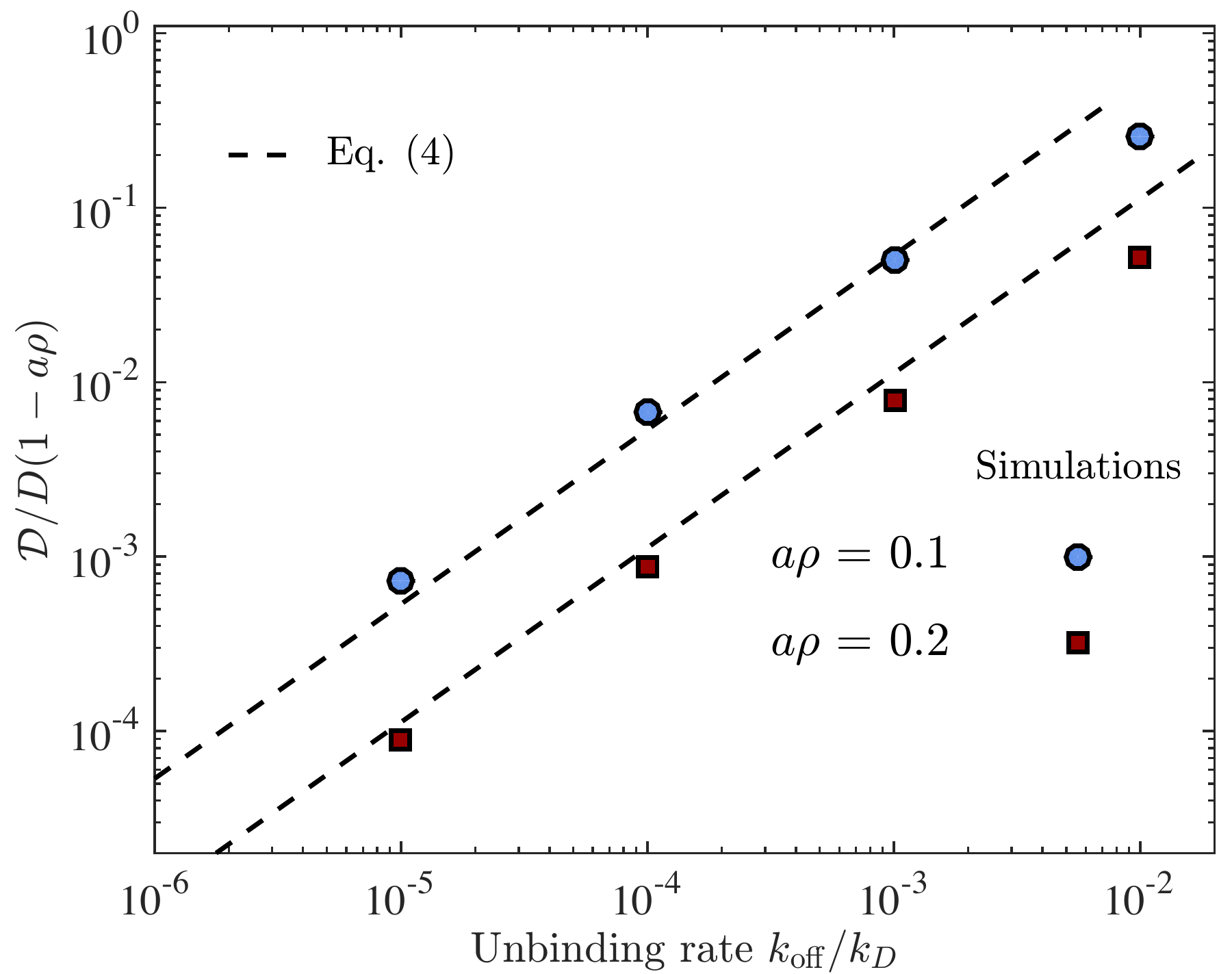}
\includegraphics[width = 0.9\columnwidth]{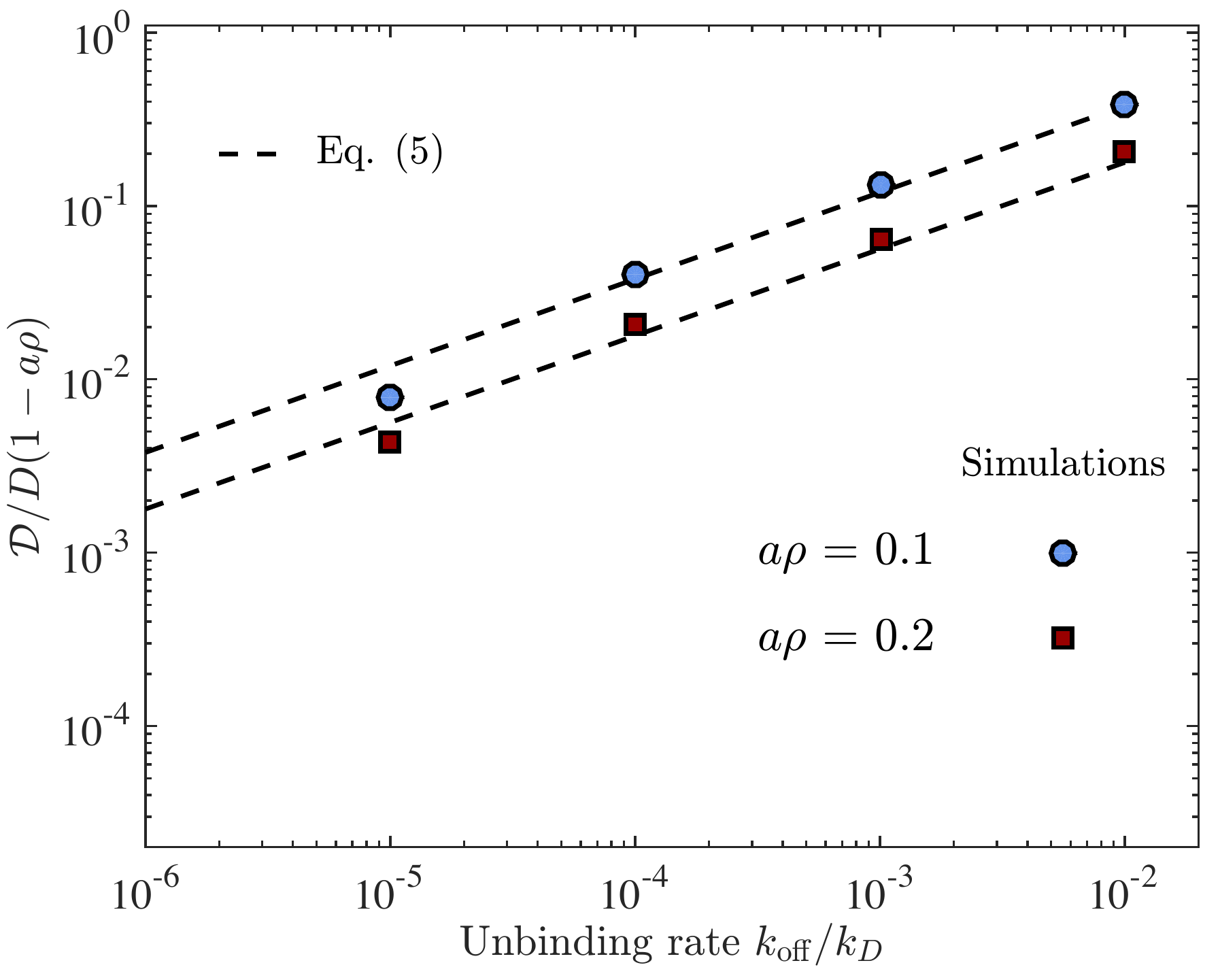}
\caption{Long time diffusion constant $\DD$ as a function the unbinding rate $\koff$, when $\koff$ is small. Symbols represent simulations for the filling fractions $a\rho=0.1$ and $a\rho=0.2$. 
%The solid lines shows the interpolation formula Eq. (\ref{eq:DD_full}), and 
The dashed lines show our predictions Eqs. (\ref{eq:D_gate})--(\ref{eq:D_sfd}). Each panel depicts: (top) immobile crowder particles with rebinding to the same location (middle) immobile crowder particles with rebinding to a random location, (bottom) diffusing crowder particles with rebinding to a random location. The data points are extracted from linear fits of Figs. \ref{fig:msd_no_diffusion}--\ref{fig:msd_diffusion} (see Appendix \ref{sec:linear_fit}). $R^2$--values from those fits are larger than 0.98.}
\label{fig:mobility_small}
\end{figure}
In Fig. \ref{fig:mobility_small}  we show how  $\DD$ depends on small $\koff$ where each panel depicts: (top) immobile crowder particles and rebinding to the same lattice site, (middle), immobile crowder particles and rebinding to a random lattice site, and (bottom)  diffusing crowder particles and rebinding to a random site. Symbols represent simulation results and dashed lines  the small $\koff$ expressions (\ref{eq:D_gate})--(\ref{eq:D_sfd}). Each case is plotted for two concentrations, $a\rho=0.1$ and $a\rho=0.2$.
% and the values for $\DD$ is extracted from Figs. \ref{fig:msd_no_diffusion} and \ref{fig:msd_diffusion} (see Appendix \ref{sec:linear_fit}). 
In order to better compare the three panels with the figures below we scaled the vertical axis  with the large $\koff$ limit $D(1-a\rho)$.  In Fig. \ref{fig:mobility_large} we show explicitly how this limit is approached.
% for different particle concentrations.

The two upper panels in Fig. \ref{fig:mobility_small}, where the crowder particles are immobile are very similar to each other. If both cases would be depicted in the same graph, the data points would practically sit on top of each other. 
%The theoretical results derived in appendix ... indicates that the difference is  $(1-a\rho)\rho^{-2}\koff$ for small k. 
For clarity,  we therefore separated the data into two figures. We see that the small $\koff$ behaviour agrees very well with the theoretical results, Eqs. (\ref{eq:D_gate}) and (\ref{eq:D_rnd}).

The lower panel depicts when crowder particles diffuse on the lattice. Their movements lead to an overall increase of $\DD$ for the tracer particle since they no longer act as static road blocks. This also changes the scaling with $\koff$ from linear in the two upper panels, to $\sqrt\koff$. The density dependence is also weaker ($1/\rho$ compared to $1/\rho^2)$. 

\subsection{Large $\koff$ behaviour}

\begin{figure}
\includegraphics[width = \columnwidth]{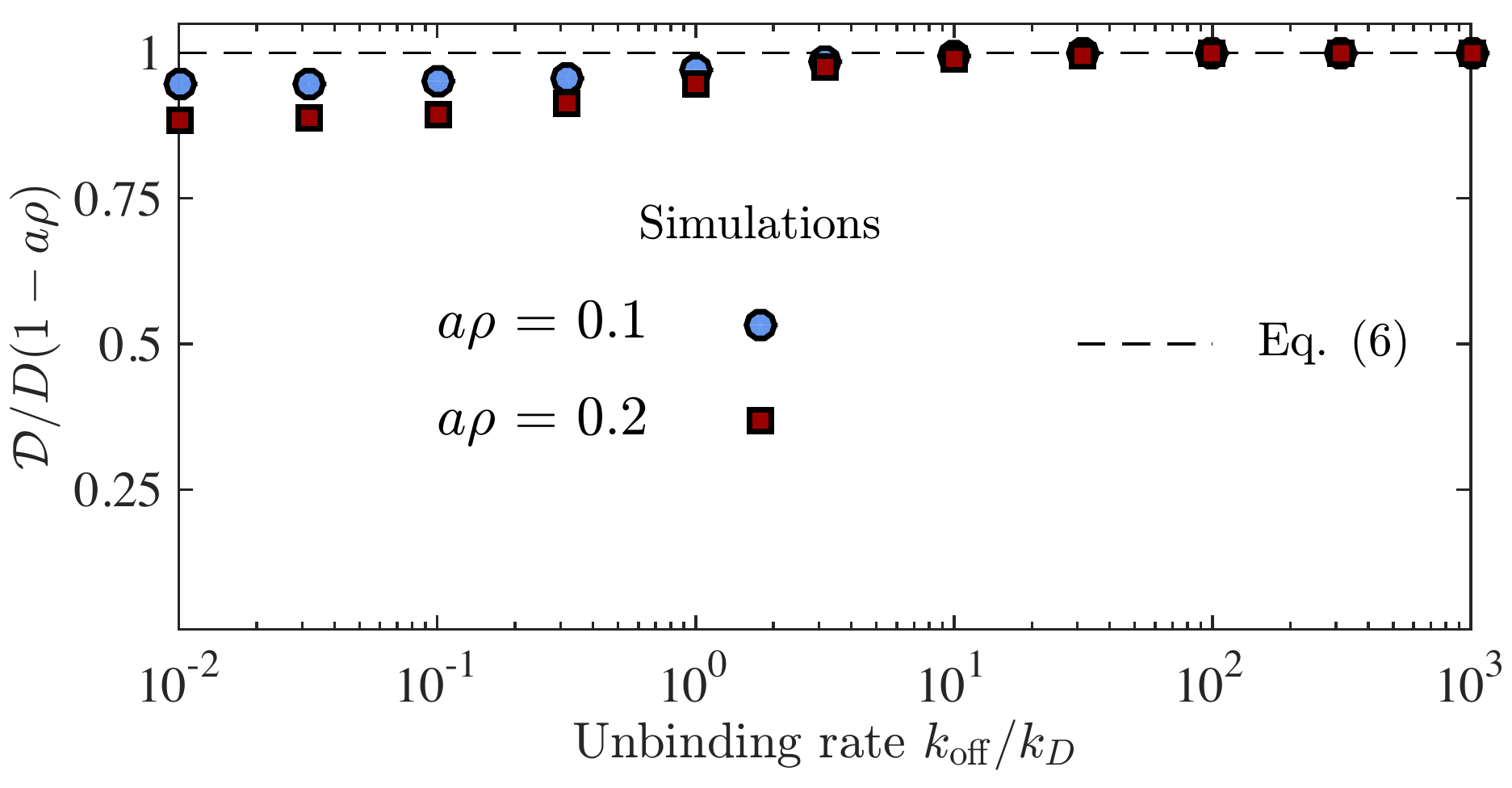}
\caption{Long time diffusion constant $\DD$ as a function the unbinding rate $\koff$, when $\koff$ is large. Simulation details are the same as in Figs. \ref{fig:msd_no_diffusion} and \ref{fig:mobility_small}.}
\label{fig:mobility_large}
\end{figure}
When $\koff$ is much larger than the diffusion rate $\kd$, we expect the mean field result Eq. (\ref{eq:D_large}) to hold. We also expect that corrections to this result becomes increasingly prominent as $\koff$ is lowered. Both are confirmed by simulations in  Fig. \ref{fig:mobility_large}, where we see that $\DD/(1-a\rho) \approx 1$ for $\koff/k_D \gtrsim1$, and $\DD/(1-a\rho)< 1$ for  $\koff/k_D< 1$. These results validate the mean field argument leading up to Eq. (\ref{eq:D_large}) for our quasi one dimensional system. The figure only shows the case where the crowder particles rebind to the same location, since the behaviour at large $\koff$ is close to identical for all rebinding modes.

\subsection{Interpolation formula}

\begin{figure}
\includegraphics[width = 0.9\columnwidth]{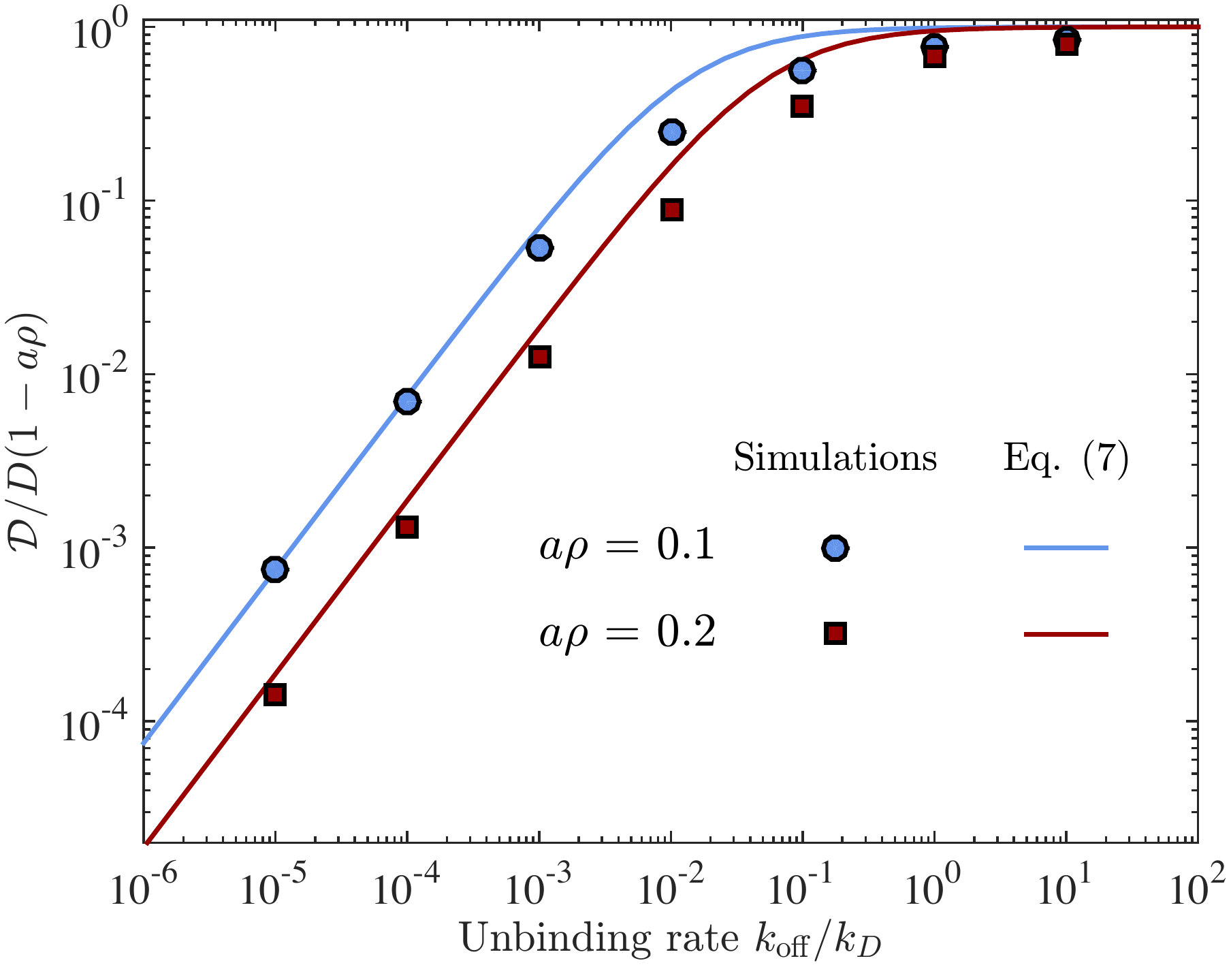}
\includegraphics[width = 0.9\columnwidth]{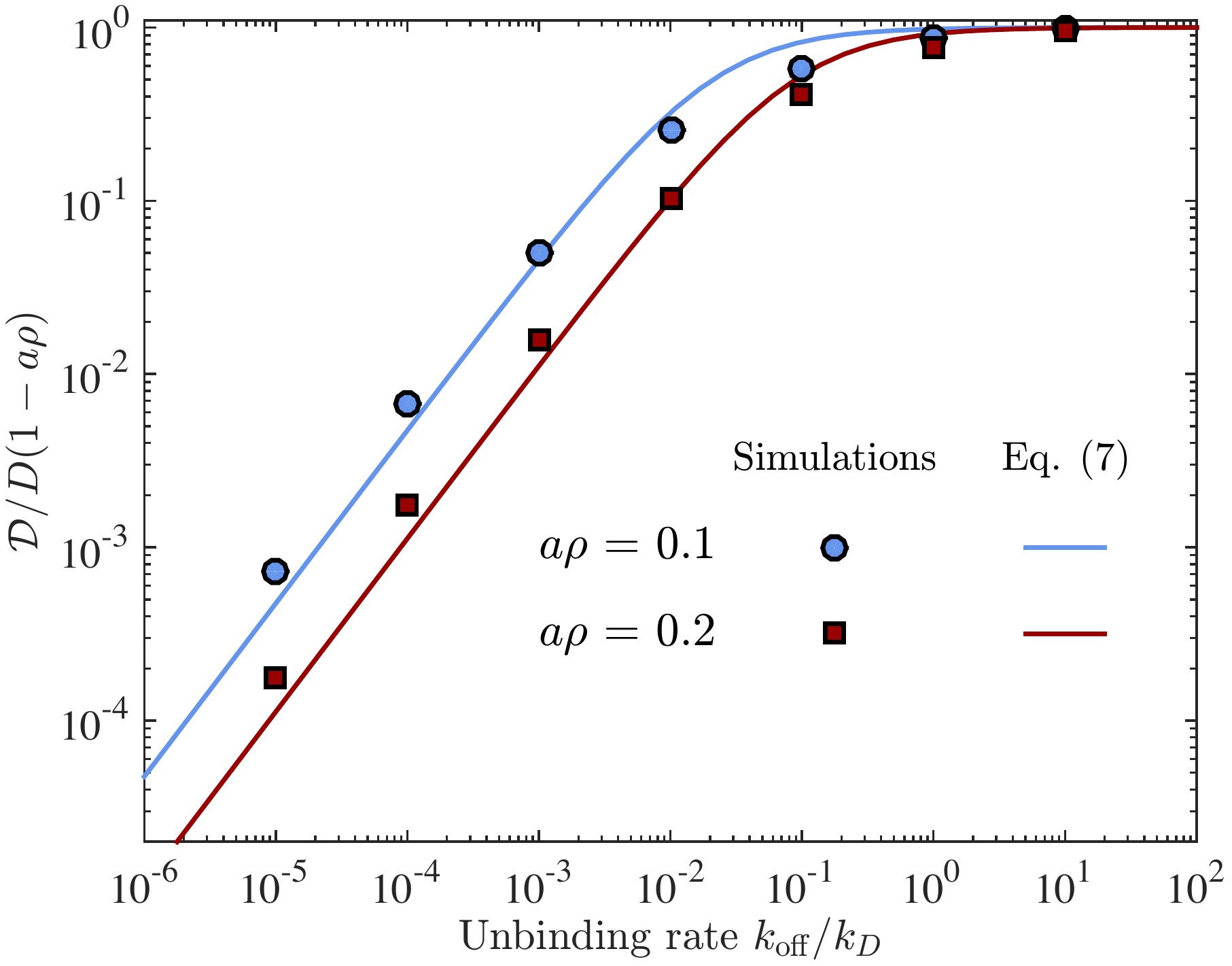}
\includegraphics[width = 0.9\columnwidth]{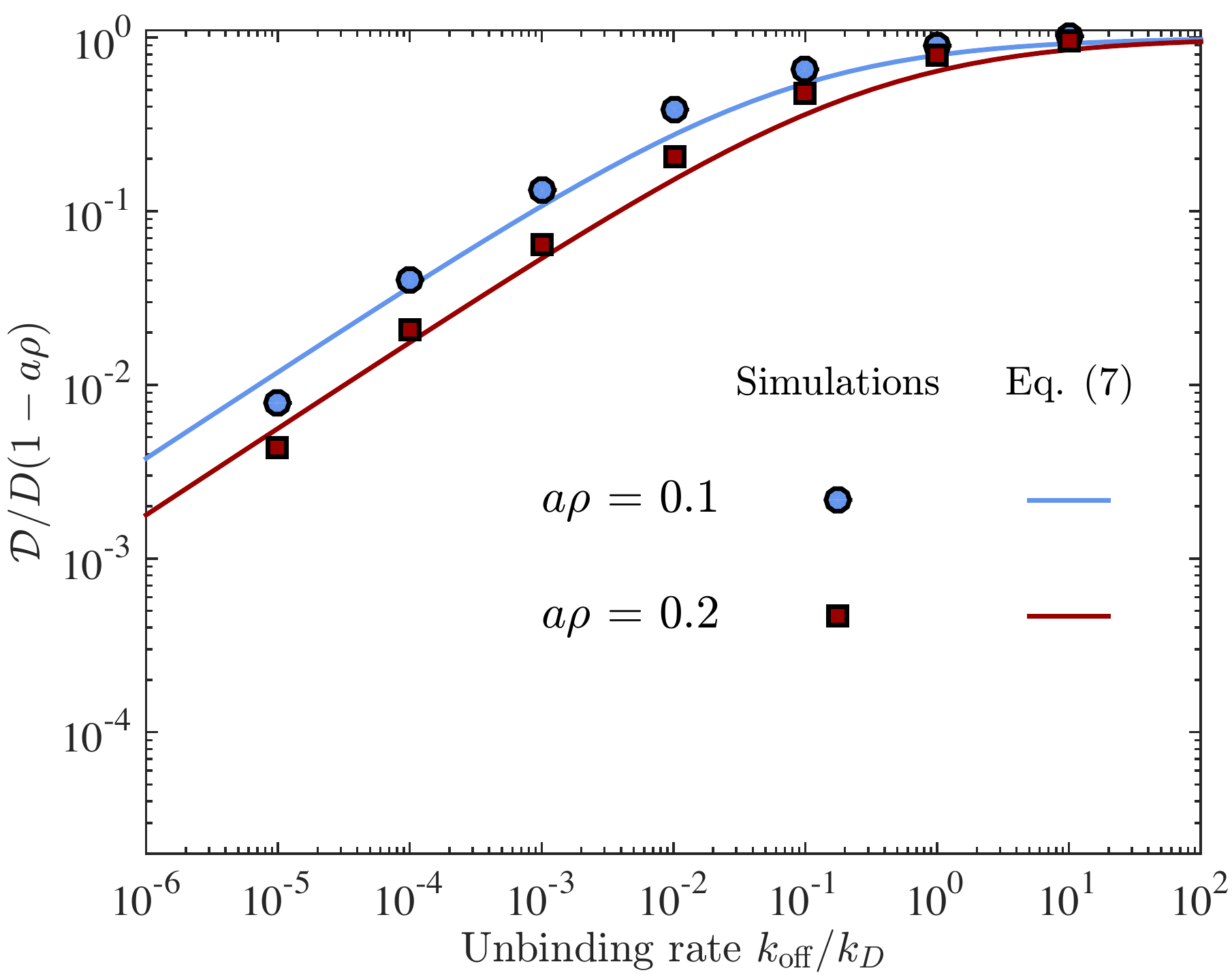}
\caption{Long time diffusion constant $\DD$ as a function of the unbinding rate $\koff$. Symbols represent simulations for two different filling fractions, $a\rho=0.1$ and $a\rho=0.2$. The solid lines shows the interpolation formula Eq. (\ref{eq:DD_full}). Each panel depicts: (top) immobile crowder particles with rebinding to the same location (middle) immobile crowder particles with rebinding to a random location, (bottom) diffusing crowder particles with rebinding to a random location. The data points are a compilation from Figs. \ref{fig:mobility_small} and \ref{fig:mobility_large}.}
\label{fig:mobility_interpol}
\end{figure}
In Sec \ref{sec:analyt} we proposed Eq. (\ref{eq:DD_full})  that ties together the small and large $\koff$ regimes. The comparison to the full range of $\koff$ is shown in Fig. \ref{fig:mobility_interpol} as solid lines (symbols are simulation results).
Just as in Fig. \ref{fig:mobility_small}, each panel shows: (top) immobile crowder particles and rebinding to the same lattice site, (middle), immobile crowder particles and rebinding to a random lattice site, and (bottom)  diffusing crowder particles and rebinding to a random site. Overall, Eq. (\ref{eq:DD_full}) is a good approximation for the whole range of $\koff$. The deviations are largest in the transition region, roughly $10^{-3}< \koff/\kd <10^{-1}$, where the maximum relative error for all curves is 79\% (top panel, $a\rho=0.2$). The relative error in the small and large $\koff$ tails is less than 7\%. 
%When crowder particles are diffusing (lower panel) the interpolation formula deviates also for small $\koff$, since our theory cannot fully capture the dynamics.

%%%%%%%%%%%%%%%%%%%%%%%%%%%%%%%%%%%%%%%%%%%%%%%%%%%%%%%
%
%		S U M M A R Y  A N D  O U T L O O K
%
%%%%%%%%%%%%%%%%%%%%%%%%%%%%%%%%%%%%%%%%%%%%%%%%%%%%%%%

\section{Summary and concluding remarks}\label{sec:conclusion}

We studied the long-time diffusion constant $\DD$ of a tracer particle in a one dimensional crowded many-particle system.  We found that $\DD$ depends strongly on the unbinding rate $\koff$ of the surrounding crowder particles and density $\rho$. For small $\koff$ we made a simple theoretical model where we deduced that $\DD \sim \koff/\rho^2$ (to first order in $1/\rho^2$) when crowder particles are immobile and only unbind/rebind to the lattice. The prefactor depends on how they rebind,  either to the same or to a random site.
% Both cases show the same dependence on $\koff$ and $\rho$, but with different prefactors. 
When they also diffuse we obtain $\DD\sim \sqrt {D \koff/\rho^2}$ (to first order in $1/\rho$), a different $\koff$-scaling than before; $D$ is the free particle diffusion constant. This means that $\DD$ is less sensitive to $\koff$ and $\rho$ when crowder particles are diffusing compared to standing still. For large $\koff$, we found that all cases agreed with the mean field result $\DD \simeq D(1-\rho a)$, independent of $\koff$. Our new expressions showed overall  good agreement with simulations.

It is interesting to see which $\koff/\kd$ regime we expect to find in the living cell. As mentioned in the introduction, residence times of DNA binding proteins vary from fractions of a second to up to an hour (unspecific binding is even shorter, $>$5 ms  \cite{elf2007probing}). To get an order of magnitude estimate of  $\koff/\kd$, let us assume that $\koff \sim 0.1$ s$^{-1}$ which lies between the  {\it in vivo} values for the LexA  and Gal4 transcription factors. One dimensional diffusion constants also have a big variation. They are in the range $D_{\rm 1D} \sim 10^5-10^7$ (bp)$^2$ s$^{-1}$  ($\approx 0.01 - 0.1$ $\mu \rm m ^2 \rm s ^{-1}$) \cite{gorman2008visualizing}, which gives
$ \kd = D_{\rm 1D}/(\rm bp)^2\sim 10^5-10^7$ s$^{-1}$. This means that $\koff/\kd~\sim 10^{-6}-10^{-8}$, which clearly indicates that $\koff\ll \kd$.

The model we studied is inspired by protein diffusion on DNA.  Our results are simple formulas for the diffusion constant of a tracer particle taking crowding and binding/unbinding dynamics into account. Although a protein is more complex than a hard-core particle,  we hope that the simplicity of our results will find its usefulness in a range of settings, in particular, single-molecule tracking experiments.

%%%%%%%%%%%%%%%%%%%%%%%%%%%%%%%%%%%%%%%%%%%%%%%%%%%%%%%%%%%%%
%
%		A C K N O W L E D G E M E N T S
%
%%%%%%%%%%%%%%%%%%%%%%%%%%%%%%%%%%%%%%%%%%%%%%%%%%%%%%%%%%%%%

\section{Acknowledgements}

LL acknowledges the Knut and Alice Wallenberg foundation  and the Swedish
Research Council (VR), grant no. 2012-4526, for
financial support. TA is grateful to VR for funding (grant no. 2009-2924).

\appendix
\section{Simple model for $\DD$ when $\koff$ is small} \label{appendix_1}

In this appendix we outline the derivation for the long time diffusion constant $\DD$ in the small $\koff$ limit that led to Eqs. (\ref{eq:D_gate})--(\ref{eq:D_sfd}). The main idea is to  calculate the typical length scale $l_0$ that the tracer travels before being hindered by a crowder particle. In terms of $l_0$, the long time diffusion constant is
\be \label{eq:DD}
\DD =  \frac{l_0^2}{2\tau },
\ee
where $\tau$ is the typical waiting time until a successful jumping event. Since the diffusion rate $\kd$ is fast, the rate limiting step for the tracer particle to move is when a flanking crowder particle unbind from the lattice. We can therefore envision the tracer particle diffusion as a single particle diffusion process on a coarse grained lattice, with lattice constant $l_0$ and jump rate $1/\tau$. 

First we address $\tau$, the average time until a successful jumping event. Imagine that the tracer particle is flanked by two crowder particles, and the time for any of them to unbind is $1/2\koff$. Now, say the that the tracer's right neighbour unbinds but the tracer  anyway tries to jump left. This jump is forbidden, and so is in fact half of all tries the tracer makes. This implies that  $\tau$ is $1/\koff$ rather than $1/2\koff$. Moreover, we also consider rebinding of crowder particles, so even if the tracer move in the direction of the unbound neighbour it may anyway be blocked by another crowder particle. We must therefore correct the jumprate with the probability that the site is vacant, that is $1-a\rho$ (in equilibrium). In summary, we estimate $\tau$ as
\be
\tau =  \frac 1 {\koff (1-a\rho)}
\ee
%

%before an unbinding event of any two of the tracer particles flanking neighbours takes place, 
%corrected by the fact that we may have a rebinding event. 
%If no rebinding would occur, the total unbinding rate would be $2\koff$ for the two neighbours. But, since we also have rebinding, the tracer may anyway be blocked and the probability that this will not happen is simply $1-a\rho$ (in equilibrium). This is similar to the mean field argument that led to Eq. (\ref{eq:D_large}). In summary, we estimate

\begin{figure}
\includegraphics[width = \columnwidth]{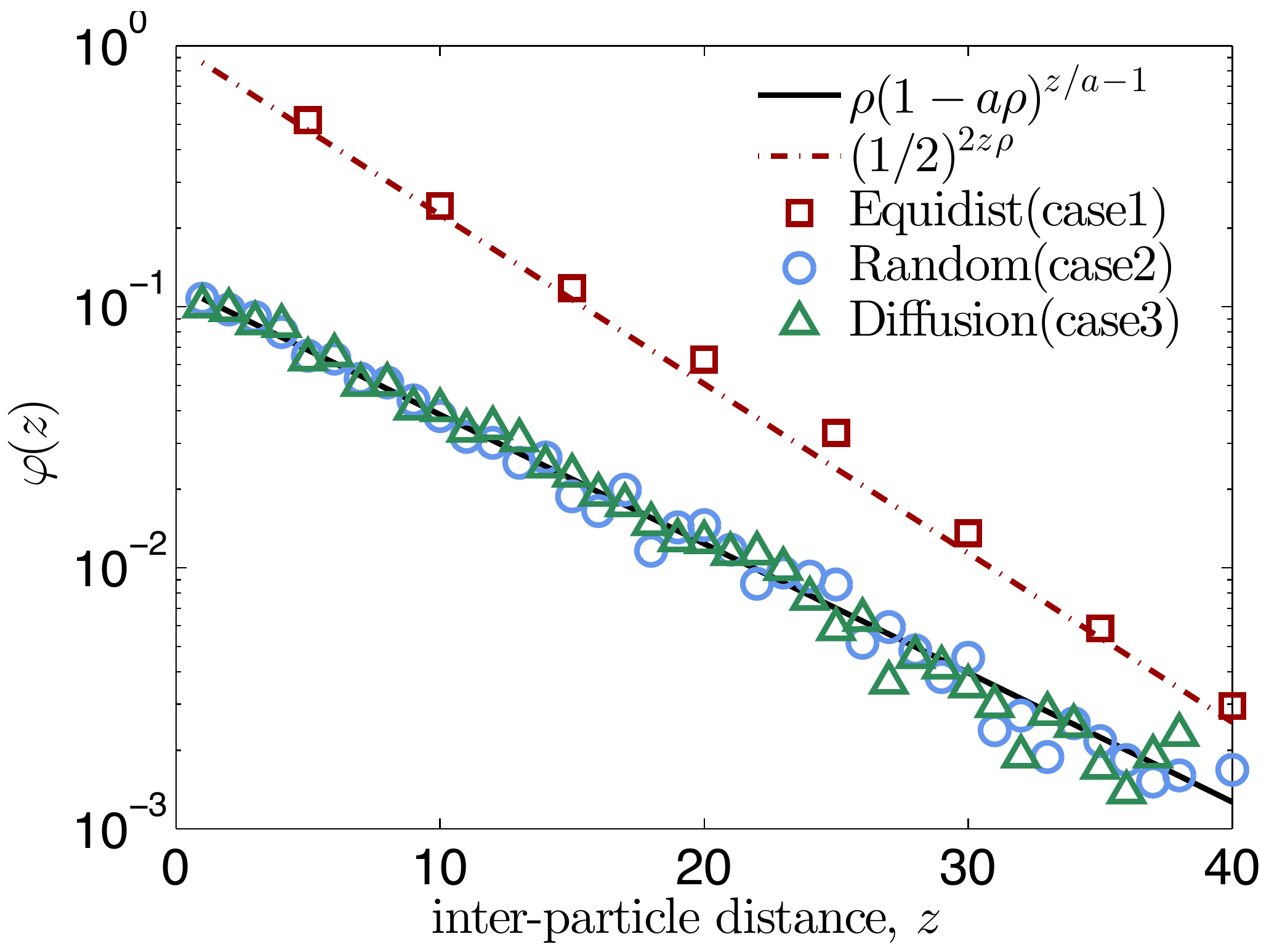}
\caption{Distribution of distances between nearest neighbours  $\varphi(z)$ in semi-log scale. Symbols depict simulation results, and dashed curves are theoretical results. The filling fraction is $a\rho=0.1$. The remaining simulation details are the same as in Figs. \ref{fig:msd_no_diffusion} and \ref{fig:msd_diffusion}.}
\label{fig:varphi}
\end{figure}
Second  we turn our attention to the coarse-grained lattice distance $l_0$. In short, we choose $l_0$ the standard deviation of the distribution of nearest-neighbour distances. Here is how we formally arrive at this result. Since it can happen that nearest and next--to--nearest neighbours are unbound simultaneously, the length that the tracer particle can move, $z$, can vary. We choose the probability distribution of $z$ to be the probability that there is a separation $z$ between two nearest neighbour particles. The distribution of $z$, $\varphi(z)$, is known (see Fig. \ref{fig:varphi}), but differs depending on the type of rebinding. If $z$ only can change in discrete steps of $\Delta$, that is  $\Delta, 2\Delta,3 \Delta...$, we can define a jump length distribution $g(l)$ for the tracer particle in the coarse grained lattice as
\be \label{eq:g(l)}
g(l) = \sum_{n=-\infty}^\infty\delta(l+n\Delta) \varphi(n\Delta),
\ee
where $\delta(x)$ is the Dirac delta function. 
%Note that $g(l)$ contains the length of the jump and its direction (left or right). 
Now we choose $l_0$ as the standard deviation of $g(l)$, and from Eq. (\ref{eq:g(l)}) one can show that
% and $\varphi(-z)=\varphi(z)$.
% In Fourier space this equation simplifies to
%
%\be \label{eq:g_k}
%g(k) = 2 \sum_{n=1}^\infty\cos(kn\Delta) \varphi(n\Delta)
%\ee
%
%where we assumed that $\varphi(-z)=\varphi(z)$. Since the width of $g(l)$ is simply $\sqrt{-g''(k=0)}$, and we take $l_0$ as the halfwidth, we obtain from Eq. (\ref{eq:g_k})
%
\be\label{eq:l_0}
l_0  = \Delta\sqrt{ \langle n^2 \rangle_\varphi}
\ee
where $\langle n^2 \rangle_\varphi = \sum_{n=1}^\infty n^2\varphi(n\Delta)$. In the subsections below, we calculate $l_0$ explicitly for the special cases: (1) immobile crowder particles placed equidistantly, (2) immobile crowder particles placed at a random distance apart from each other, and (3) diffusing crowder particles.

\subsection{Case 1: Immobile crowder particles placed equidistantly}

In the simulations, the crowder particles sit equidistantly, and unbind and rebind with rates $\koff$ and $\kon$, respectively. In order to make sure that the average density $\rho$ is constant over time, we choose $\kon=\koff$, and work with $2m$ particles in the whole system (lattice + surrounding bulk). This means that $m$ particles will on average be on the lattice and the density is $\rho = m/(aN)$, where $N$ is the number of lattice sites, and $a$ the lattice constant. The smallest separation between two crowder particles in this setup becomes $\Delta=1/(2\rho)$, and increase discrete steps of $\Delta$:
\be
 \frac 1 {2\rho}, \ \frac 1 {\rho}, \ \frac 3 2 \frac 1 {\rho},\ \ldots
\ee
Each one of these lengths has a different probability, and the distribution of nearest neighbour distances is 
\be
\varphi\left( \frac{n}{2\rho}\right) = \left\{\begin{array}{ll}
\frac 1 2 \left(\frac 1 2\right)^{|n|},    & n = \pm 1, \pm 2, \ldots \\
0,	& n=0,
\end{array}
\right. 
\ee
which agrees well with simulations (Fig. \ref{fig:varphi}). 
%It is now a straightforward matter to calculate $l_0$. 
Using that $\langle n^2\rangle_\varphi=6$ in Eq. (\ref{eq:l_0}) gives
\be
l_0 =  \frac 1 \rho \sqrt{\frac 3 2}.
\ee
%
%and $\DD = (3/4) \koff/\rho^2$ when $\tau = 1/(2\koff)$.

\subsection{Case 2: Immobile crowder particles with rebinding to random locations}

In this case the crowder particles leave the lattice and return to a random vacant lattice site. This means that the smallest separation is  the lattice distance of the original lattice, $\Delta=a$, and distances are in steps of a:
\be
a,\ 2a,\ 3a,\ \ldots
\ee
The inter-particle distance distribution in this case is 
\be
\varphi(na) = \left\{\begin{array}{ll}
\frac 1 2 \frac {a \rho}{1-a\rho} (1-a\rho)^{|n|},    & n = \pm 1, \pm 2, \ldots \\
0,	& n=0
\end{array}
\right. 
\ee
which is corroborated by simulations in Fig. \ref{fig:varphi}. In the continuum limit (small $a$), the distribution becomes exponential $\varphi(na)\sim e^{-|n|a\rho}$. Using that  $\langle n^2\rangle_\varphi= (2-a\rho)/(2 a^2\rho^2)$, we obtain
\be
l_0 =  \frac 1 \rho  \sqrt{\frac{2-a\rho}{2}}.
\ee
%
%which gives $\DD=(2-a\rho)/(4\rho^2)$ if $\tau = 1/(2\koff)$.

\subsection{Case 3: Diffusing crowder particles with rebinding to random locations}

Here all particles diffuse which drastically changes the situation. The main difference is that the tracer does not get stuck between two flanking road blocks since they also move. However, we know from simulations that the MSD for the tracer is in the long time limit proportional to $\DD t$ (Fig. \ref{fig:msd_diffusion}), which is a direct manifestation that the no-passing condition is violated (otherwise we would have had MSD $\sim \sqrt t$). Altogether, this implies that there is length scale for the coarse grained lattice and a time scale associated with a jumping event. 

For this case we cannot use $\varphi(z)$ to estimate $l_0$ since $\varphi(z)$ is the same as when crowder particles are immobile (see Fig. \ref{fig:varphi}, $\bigcirc$ and $\bigtriangleup$), and gives the wrong result for $\DD$. The reason is that inter-particle distances fluctuate at the same rate as the tracer is diffusing, and those fluctuations increase $\DD$. In fact, even if $\koff=0$, the tracer particles still can move across the system, although slowly. We estimate $l_0$ as the distance the tracer particle explores in a time $\tau$, that is 
\be
\l_0 = \sqrt{\left\langle x^2( t= \tau) \right\rangle} = \left( \frac{4D}{\pi\koff} \frac{1-a\rho}{\rho^2}\right)^{1/4}\\
\ee

Interestingly, $l_0$ depends on $\koff$ and not only $\rho$ as in the previous cases. This changes the scaling of $\koff$ in  $\DD$ from linear in Cases 1 and 2  to $\sqrt{\koff}$ for this case. This can also be understood from the following simple argument. The curve for  $\msd$ is continuous for all times, and at some time the dynamics changes behaviour from single-file ($\sim\sqrt t$) to regular diffusion ($\sim t $). This occurs around $t\approx\tau$, which implies
\be
2\DD t \Big|_{t=\tau}\approx \left. \sqrt{\frac{4Dt}{\pi \rho^2 }} \  \right|_{t=\tau}.
\ee
This yields $\DD  \propto \sqrt{\koff}$.

%%%%%%%%%%%%%%%%%%%%%%%%%%%%%%%%%
\section{Extraction of the long time diffusion constant} 
\label{sec:linear_fit}

\begin{figure}
\includegraphics[width = \columnwidth]{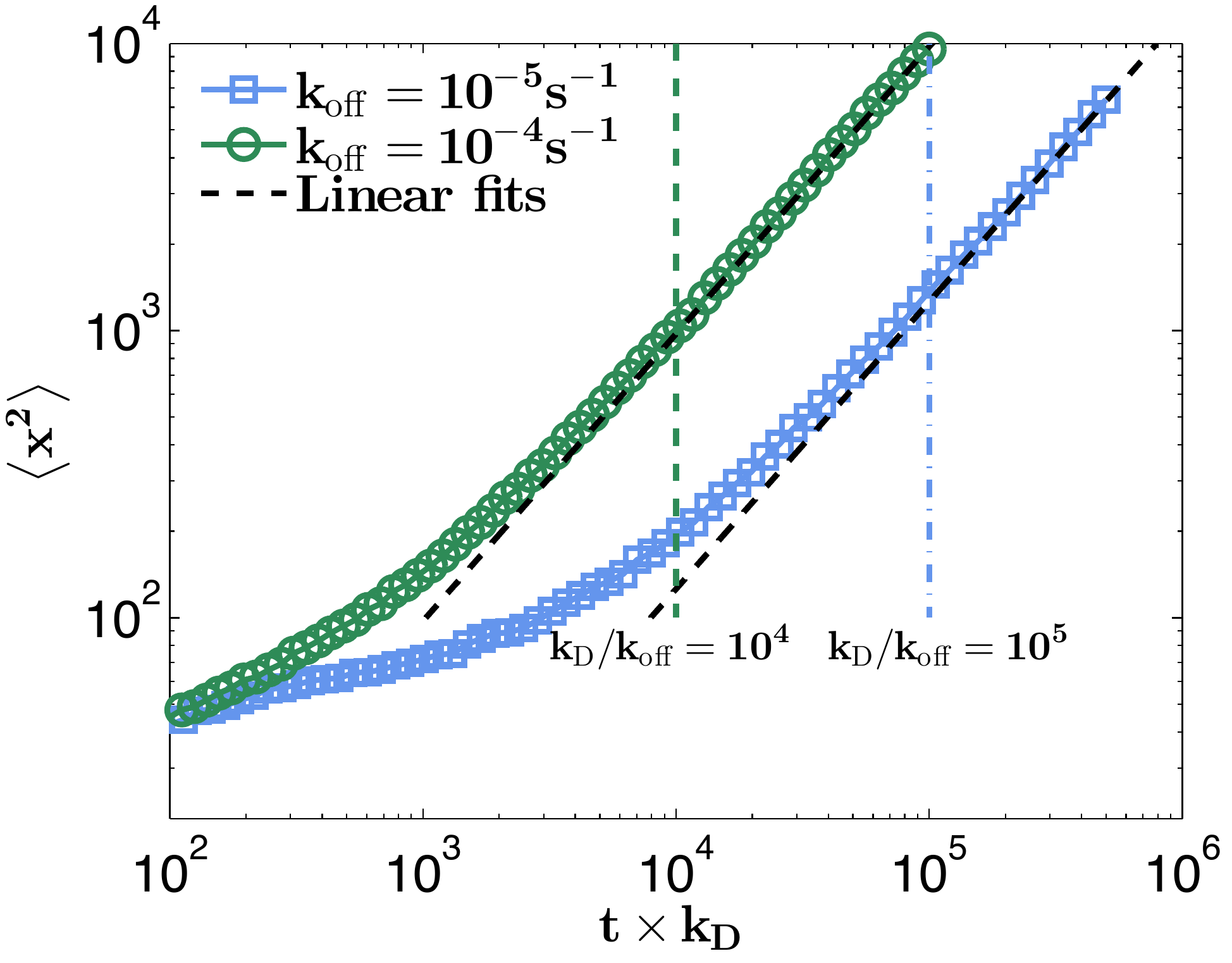}
\caption{MSD of the tracer particle as a function of time for two different unbinding rates. The dashed black lines are linear fits of the MSD curves for times $t\geq\toff$. The simulation details are the same as in Fig. \ref{fig:msd_no_diffusion}.}
\label{fig:msd_fit}
\end{figure}
The way we determine $\DD$ from our  MSD simulations, is illustrated in Fig.  \ref{fig:msd_fit}. First,  $\toff$ is the approximate time at which the MSD becomes linear (shown as vertical dashed-dotted lines). Second, we make a linear regression of the MSD curve starting from that point, and obtain the slope which equals $2 \DD$. The resulting fits are shown as dashed lines.

%%%%%%%%%%%%%%%%%%%%%%%%%%%%%%%%%
\section{Numerical implementation} 
\label{sec:numerics}

The model (Fig. \ref{fig:model})  is implemented  using the Gillespie algorithm \cite{gillespie1976general}. 
%
%Each particle on the lattice is given rates corresponding to its allowed actions: a rate of jumping left or right ($\kd$) and a rate for unbinding ($\koff$). The no-passing condition between particles is implemented by setting the jump rate to zero in the direction of a flanking particle. The tracer particle is different from all others since the binding and unbinding rates are put to zero ($\koff=\kon=0$) throughout the simulation. Its diffusive motion is otherwise the same.  
%
The majority of the details of the implementation has been explained elsewhere \cite{forsling2014non}, but below we point out some key differences.

We keep track of the unbound crowder particles in the bulk in order to have the option to rebind them at the location they detached from. In practice we use two lattices, one of which represents the bulk. The filling fraction is maintained at the level we want by setting $\kon=\koff$, and then let the systems equilibrate such that half the number of crowder particles sit in the bulk and the other half on the lattice we are interested in. This representation helpful to investigate all sorts of binding modes especially rebinding to the same location. In  Ref. \cite{forsling2014non} the bulk served as an infinite particle reservoir and the concentration on the lattice was tuned via detailed balance (rebinding always occurred to a randomly chosen site). Here on the other hand, the bulk has a finite size and cannot be seen as a strict particle reservoir. However, since we use about 500 particles, fluctuations around the filling fraction $a\rho$ are so small that we rarely (if ever) deplete the bulk. This means that we have approximately a grand canonical ensemble. 

\end{document}